\documentclass{article}
\usepackage{spconf,amsmath,graphicx}
\usepackage{amssymb}
\usepackage{booktabs}
\usepackage{tabularx}
\usepackage{tikz}
\usepackage{amsfonts}
\usepackage{hhline}
\usepackage{xfrac}

\usepackage{siunitx}
\usepackage{pdfpages}
\usepackage{pgfplots}
\usepackage{subfig}
\usepackage{algorithmicx}
\usepackage{hyperref}
\usepackage{algorithm}
\usepackage[inkscapeformat=png]{svg}
 \usepackage{relsize}
\usepackage[noend]{algpseudocode}
\algrenewcommand\algorithmicprocedure{\texttt{procedure}}
\algrenewcommand\alglinenumber[1]{\tiny #1:}
\makeatletter
\renewcommand{\ALG@beginalgorithmic}{\small}
\makeatother
\usetikzlibrary{spy,calc}
\usepackage{hyperref}
\usepackage{subfig}

\title{PUSHING THE LIMITS OF THE WIENER FILTER IN IMAGE DENOISING}
\name{Clément Bled, François Pitié\thanks{This research is supported by Science Foundation Ireland in the ADAPT Centre (Grant 13/RC/2106) (www.
adaptcentre.ie) at Trinity College Dublin.}}
\address{Department of Electronic and Electrical Engineering\\
Trinity College Dublin
}

\usepackage{xcolor}

\begin{document}
\ninept
\setlength{\textfloatsep}{10pt}
\maketitle
\begin{abstract}
As modern image denoiser networks have grown in size, their reported performance in popular real noise benchmarks such as DND and SIDD have now long outperformed classic non-deep learning denoisers such as Wiener and Wavelet-based methods. In this paper, we propose to revisit the Wiener filter and re-assess its  potential performance. We show that carefully considering the implementation of the Wiener filter can yield similar performance to popular networks such as DnCNN.
\end{abstract}
\section{Introduction}
\label{sec:intro}

Despite advances in camera sensor technologies, image denoising still remains a key part in many application
pipelines. Denoisers, such as the Wiener~\cite{pratt1972generalized, king1983wiener, giger1984investigation, benesty2010study}, Wavelet~\cite{mallat1989theory, combettes2004wavelet, malfait1997wavelet}, and BM3D \cite{dabov2007image} filters have now been significantly outpaced by neural networks, whose architectures can easily be trained~\cite{zhang2017beyond, zhang2018ffdnet} to achieve blind denoising, without the need to provide estimates of the noise profile, and which can also be efficiently adapted to operate beyond Gaussian white noise~\cite{guo2019toward}. On the real-noise DND benchmark~\cite{plotz2017benchmarking}, it can be observed that the performance increased from about 35dB on non-CNN state-of-the-art denoisers (BM3D), to about 37dB using early CNNs (DnCNN~\cite{zhang2017beyond}, FFDNet~\cite{zhang2018ffdnet}), and up to 40+\,dB using the largest, transformer-based networks. 

Part of recent performance gains may be attributed to new methods of synthesising pseudo-real, signal-dependent noise for training datasets \cite{guo2019toward}, as well as the adoption of UNet \cite{Liu2019MWCNN, zamir2021multi} and vision-transformer backbone architectures (Restormer~\cite{zamir2022restormer}, SwirIR~\cite{liang2021swinir}). The network sizes have, however, also steadily increased, with parameter counts now reaching 44 million, where earlier architectures such as DnCNN only proposed 0.6 million parameters. The latest neural networks have thus become slower and impractical to run on machines without high VRAM and high core count GPUs.  


In this paper, we propose to revisit the classic Wiener filter and show how its baseline implementation (see section~\ref{sec:background}) can be improved by a number of careful adjustments (see section~\ref{sec:method}) to bring its performance in line with more recent deep learning denoisers. In particular, our ablation studies (section~\ref{sec:results}) show that, by using tiny ancillary networks to estimate some of the Wiener parameters, we can propose a hardware-friendly Wiener denoiser that can perform on par with DnCNN, one of the most popular CNN-based denoisers.





\section{Background} 
\label{sec:background}

Given a noisy signal $y$, composed of the original, unknown signal $x$, and additive noise $n$, $y = x + n$;
the Wiener filter \cite{wiener1949extrapolation} defines a  linear, minimum mean square error (MMSE) optimal filter. Assuming that the image and noise signal are second-order stationary and  decorrelated,  the optimal IIR Wiener filter is given by the following transfer function $H(\omega_1,\omega_2)$: 
\vspace{-0.5em}
\begin{equation}
    H(\omega_1,\omega_2) = \frac{ S_{xx}(\omega_1, \omega_2) }{ S_{yy}(\omega_1, \omega_2)}\,,
\end{equation}
where $S_{yy}(\omega_1,\omega_2)$ and $S_{xx}$ are the power spectrum densities at spatial frequencies $\omega_1,\omega_2$ for the input signal $y$ and original signal $x$. 
In practice, the PSD of the unknown, clean signal is estimated as $S_{xx} \approx S_{yy} - S_{nn}$, which leads to the following \textit{coring} function:
\vspace{-0.1em}
\begin{equation}\label{eqn: wiener_coring}
    \hat{S}_{xx}(\omega_1,\omega_2) = \max( S_{yy}(\omega_1, \omega_2) - S_{nn}(\omega_1, \omega_2), 0).
\end{equation}


If the noise is Additive White Gaussian (AWGN), the PSD is a constant $P_{nn} \propto \sigma^2 $, where $\sigma$ is the standard deviation (STD) of the noise. A correcting factor $1.4 \times \sigma$ is typically applied to better remove the noise.

As images are not stationary signals, the input image needs to be broken into overlapping blocks (eg. 32\texttimes32), followed by a windowing function (eg. half-cosine) before applying the FFT. The blocks are typically overlapped by half a block size, in a 2:1 overlap configuration~\cite{anilbook}. In this configuration, using the same half-cosine window for both analysis and synthesis yields a net effective signal gain of 1 when the overlap blocks are summed. This baseline implementation is summarised in Alg.~\ref{alg:WienerBaseline} (the symbols $\odot$ and $\oslash$ denote element-wise multiplication and divisions in the blocks).

The Wiener filter can produce characteristic ringing artefacts around edges and textures when the parameters (eg. noise PSD) are incorrect. This is a well-observed effect of Fourier series compression known as Gibb's phenomenon~\cite{hewitt1979gibbs}.

{ \setlength{\textfloatsep}{5pt}
\begin{algorithm}[tb]

\small 
\caption{Baseline Wiener for a Raised Cosine Window\label{alg:WienerBaseline}
}
\begin{algorithmic}[1]
    \Require Noisy image, $y$, noise STD $\sigma$, Block Size
    \State $w(h,k) \gets RaisedCosine(h,k)$ \Comment{windowing definition}
    \ForAll {blocks ${\bf y}$ in image $y$, for stride=BlockSize/2}
    \State $\bar{y} \gets \mathrm{mean}({\bf y})$ \Comment{predicts block mean}
    \State ${\bf y}_w \gets ({\bf y} - \bar{y}) \odot {\bf w}$ \Comment{windowing}
    \State ${\bf Y} \gets FFTn({\bf y}_{w})$ 
    \State ${\bf P}_{yy} \gets {\bf Y} \odot {\bf Y}^*$
    \State ${\bf P}_{nn} \gets \hat{\sigma}^2 \| {\bf w} \|^2$
    \State ${\bf P}_{xx} \gets \max({\bf P}_{yy} - P_{nn}, 0)$ \Comment{coring}
    \State $\hat{\bf x}_{w} \gets iFFTn({\bf Y} \odot {\bf P}_{xx} \oslash {\bf P}_{yy}) + \bar{y}  {\bf w}$
    \State ${\hat x} \gets \mathrm{overlap\_add}({\bf w} \odot \hat{\bf x}_{w})$ \Comment{combine blocks}
    \EndFor
\end{algorithmic}
\end{algorithm}
}

\section{A Multi-Scale, Overlapping Wiener Filter}
\label{sec:method}

We propose here to optimise some aspects of the baseline Wiener.


\subsection{Block Processing and Windowing}\label{sec:windowopt}

{\noindent \bf Gaussian Window.} In the baseline implementation, the half-cosine windowing function is used for both FFT analysis and 2D interpolation of the filtered blocks. We first propose that a Gaussian window is more appropriate as it is isotropic. Results from section~\ref{sec:results} show a+\,0.1\,dB improvement. 

\vspace{.25em}
{\noindent \bf Finer Overlaps.} Instead of using half-bock strides between blocks, we propose that quarter-block or finer overlaps can yield improved results (+\,0.5\,dB). To achieve this, we propose in Alg.~\ref{alg:ProposedWiener} to slightly modify the baseline Wiener implementation so that we can parse images with denser block overlaps. To make sure that the effective gain remains 1, we need to keep track of the windowing with a normalisation mask $w_{all}$ (note that this is also required by the use of a Gaussian window instead of a half-cosine one).

\vspace{.25em}
{\noindent \bf Multi-Scale.} Interestingly, this implementation also allows us to extend the processing to blocks of different sizes. The idea here is that, by combining blocks from 8\texttimes8 to 64\texttimes64, we re-enforce the assumption that the decorrelation between the noise and the signal should happen at all scales. This multi-scale overlap yields consistent gains of about +\,0.3\,dB.

\subsection{Per-Block Noise Estimation}

One issue when working with real ISO camera noise, is that the assumption that the signal and noise are uncorrelated is no longer true. In fact, ISO camera noise is better modelled with a Poissonian-Gaussian signal-dependent distribution. The noise variance is inversely proportional to the irradiance incident on the sensor, with dark areas having greater noise variances than bright areas. 

We propose to mitigate this by measuring the noise STD on a per-channel and per-block basis, instead of the typical per-image basis. As this cannot be practically done by a user, we propose to train a lightweight neural network that can predict a per-pixel noise STD. This network will thus effectively transform the Wiener filter into a \textit{blind} denoiser.

With a focus on minimising network size, we will study three different network depths: 2, 4 and 6 layers; with three different layer sizes: 16, 32 and 64 channels; making for a total of 9 networks. Each layer consists of a 2D convolution, a batch-normalisation and a ReLU activation function (see  supplementary material~\footnote{\url{https://github.com/MrBled/ICICP_2023_Wiener}}).

The network is trained in two stages. First, the network is trained with a $L_1$ loss on a synthetic dataset, where the  ground-truth noise standard deviation maps are estimated from the generation, for  each clean image of the dataset, of 12 instances of signal-dependent synthetic noise, using the noise  model proposed in CDBNet~\cite{guo2019toward}. 

The model is then integrated into the Wiener filter and fine-tuned end-to-end using the $L_1$ loss between the Wiener output and the ground truth image. The Wiener filter is implemented in a differentiable manner to allow for this. In this second stage, both synthetic and real data can be used.

\begin{figure}
    \centering
  \includegraphics[width=.725\linewidth]{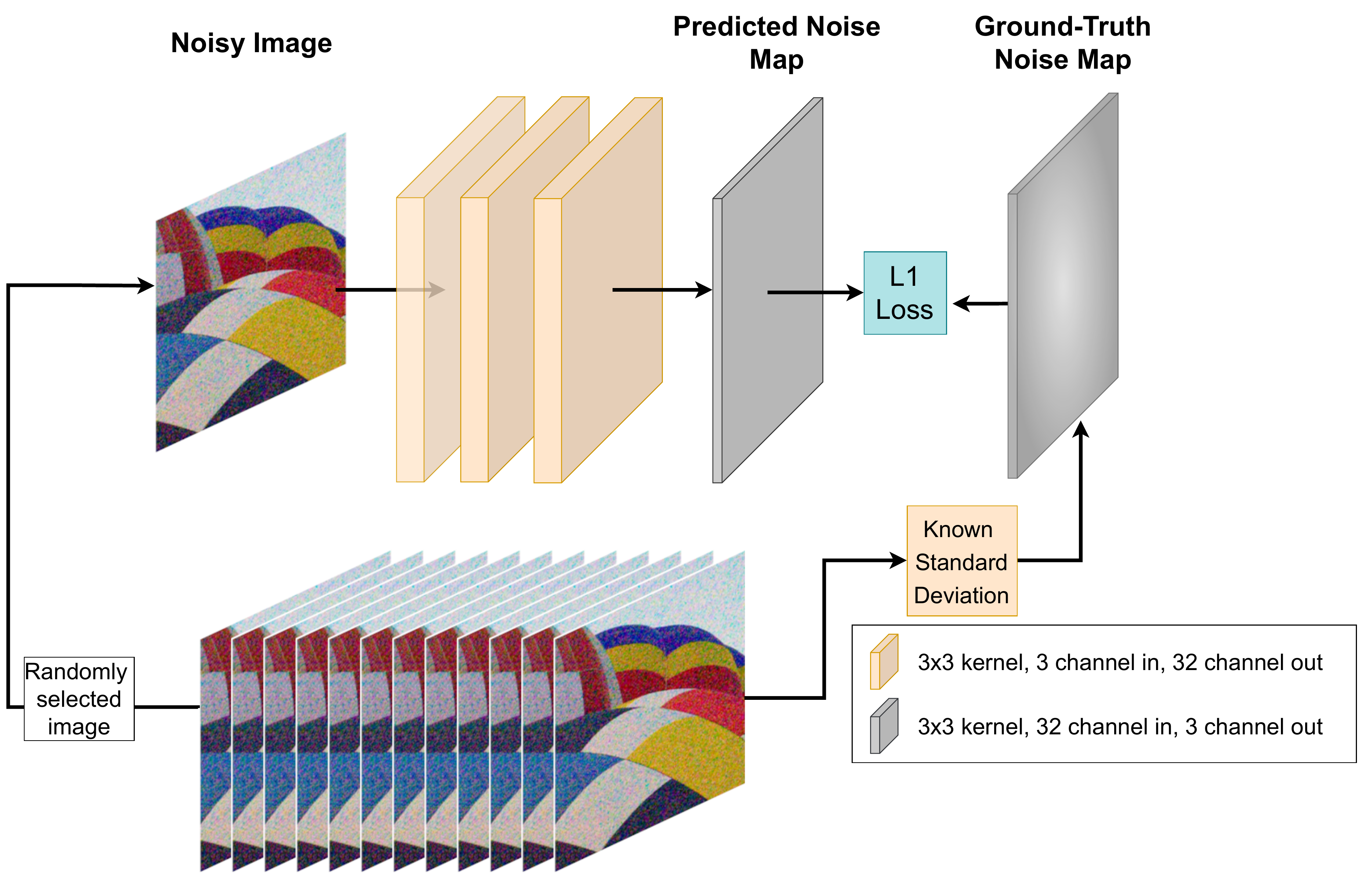}
  \caption{Training pipeline for Stage 1 of the standard deviation prediction network. In stage 2, the network is deployed and trained alongside the Wiener filter.}
  \label{fig:std_pred}
\end{figure}

\subsection{Block Mean Prediction}\label{sec:mean}
Before FFT analysis in the Wiener filter, the DC offset is subtracted from the 2D signal (see section~\ref{sec:background}). In the case of heavy noise, the Wiener coring will suppress most of the signal, only leaving this DC offset. Estimating a DC value as close to the ground-truth DC value is thus critical for the overall performance. The issue is that real noise is not necessarily zero-mean over the block. This means that the block average of the noisy signal is not a good estimate of the block average of the clean signal. This is especially true for dark areas, where the noise creates a positive shift as pixel values are never negative.

We propose here to study two solutions. Firstly, we replace the block average with the block median. The median will be more robust in dark areas, as outlier values will not bias the DC estimate. 

The second approach we want to investigate, is to train a dedicated ancillary CNN. While the variance-predicting CNN was trained using a shallow network, predicting the mean of a noisy image block requires a greater receptive field. As such, we customise a UNet, by reducing the trainable parameters to a fraction of the original network, replacing skip connection concatenations with summations and removing a downsampling layer. 

\subsection{Coring Refinement Network}\label{sec:coring_NN_method}
Finally, we explore the possibility of refining the default coring function with a small frequency-domain CNN. After applying coring on all blocks at a particular block size, 
we can collate all the values for $H(\omega_1, \omega_2)$ into a single 4D tensor of size $M_x$\texttimes$M_y$\texttimes$N$\texttimes$N$, where $M_x$ and $M_y$ are the numbers of blocks in each direction, and $N$ is the block size. We aim here to train a CNN to fine-tune this tensor. 

\begin{figure}[t]
  \includegraphics[width=\linewidth]{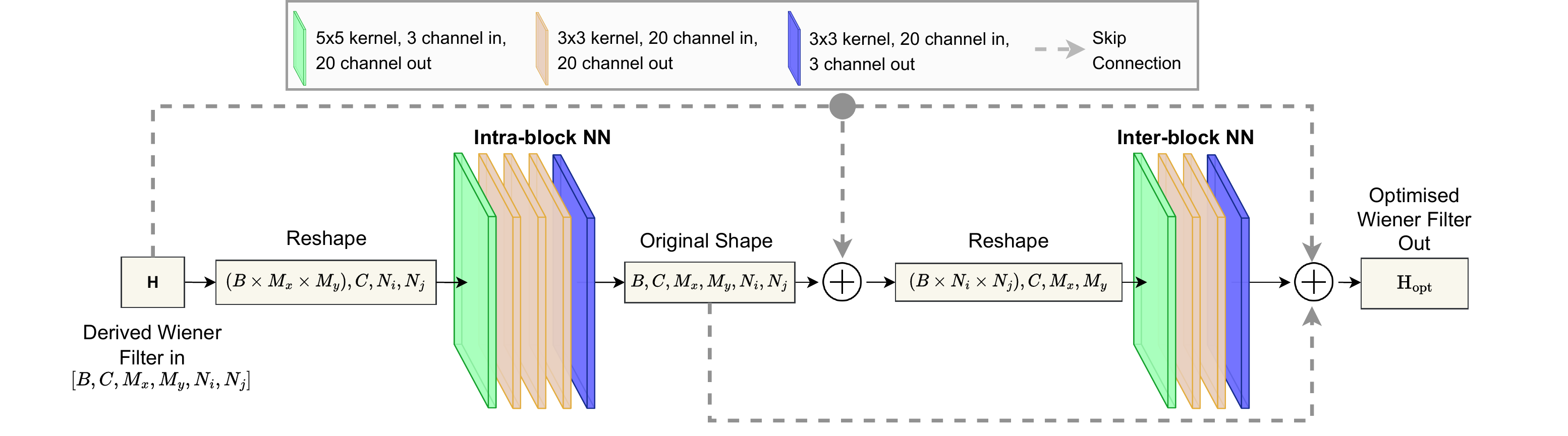}
  \caption{The Coring Network Architecture used to optimise the initial prediction of the coring function $H(\omega_1, \omega_2)$. }
  \label{fig:coring_cnn}
\end{figure}

We could potentially employ 4D convolutions filters, by combining across all frequencies of each of the neighbouring blocks, but to keep the computations reasonable, we separate the convolutions into a block of 2D convolutions operating on the frequencies within a single block (\textit{intra-block NN}), and then on a block of 2D convolutions operating on frequencies across the neighbouring blocks (\textit{inter-block NN)}. Skip-connections are added between each stage of the network (see Fig.\ref{fig:coring_cnn}). The network is kept simple, as we use the same 2D convolution, batch normalisation, ReLU activation structure as in our STD estimation network. We have 20 channels per layer, with 5 layers in stage 1, and 4 layers in stage 2. The final number of trainable parameters in the network is 22,506. As with the STD estimation network, the Wiener filter is integrated into the network, and the loss is taken as the $L_1$ loss for the resulting estimated denoised image.

\begin{algorithm}[tb]
\caption{Proposed Wiener Filter Implementation}
\label{alg:ProposedWiener}
\begin{algorithmic}[1]
    \Require Noisy image, $y$ 
    \State $\hat{\sigma} \gets \mathrm{CNN}_{\sigma}(y)$ \Comment{noise STD prediction}
    \State $w(h,k) \gets \exp\left( - \alpha (h^2+k^2) \right)$ \Comment{windowing definition}
    \ForAll {block sizes $\in [8, 16, 32, 64, 96] $}
    \ForAll {blocks ${\bf y}$ in image $y$, for stride=BlockSize/4}
    \State $\bar{y} \gets \mathrm{median}({\bf y})$ \Comment{predicts block mean}
    \State ${\bf y}_w \gets ({\bf y} - \bar{y}) \odot {\bf w}$ \Comment{windowing}
    \State ${\bf Y} \gets FFTn({\bf y}_{w})$ 
    \State ${\bf P}_{yy} \gets {\bf Y} \odot {\bf Y}^*$
    \State ${\bf P}_{nn} \gets \hat{\sigma}^2 \| {\bf w} \|^2$
    \State ${\bf P}_{xx} \gets \max({\bf P}_{yy} - P_{nn}, 0)$ \Comment{coring}
    \State $\hat{\bf x}_{w} \gets iFFTn({\bf Y} \odot {\bf P}_{xx} \oslash {\bf P}_{yy}) + \bar{y}$
    \State $x_{all} \gets \mathrm{overlap\_add}({\bf w} \odot \hat{\bf x}_{w})$ \Comment{combine   image blocks}
    \State $w_{all} \gets \mathrm{overlap\_add}({\bf w} \odot {\bf w})$ \Comment{combine all windows}
    \EndFor
    \EndFor
    \State $\hat{x} \gets \frac{x_{all}}{w_{all}}$
\end{algorithmic}
\end{algorithm}

\section{Experimental Results}
\subsection{Evaluation Datasets}\label{sec:results}
All of our results are evaluated on 50 128x128 patches from the 4k Smartphone Image Denoising Dataset (SIDD)~\cite{abdelhamed2018high}. During the period of research, both online benchmarking services: SIDD and DND were offline. We measure our performance using peak-signal-to-noise ratio (PSNR) in decibels (dB).

\subsection{Window-Based Optimisations Results}
\textbf{Window Overlap.} We first evaluate the effect of increasing the overlap between blocks. Using a half-cosine window and a block size of 38\texttimes38, we increased the block overlap from the default \sfrac{1}{2}-block overlap, to \sfrac{1}{7} of a block stride overlap. 
Results in Table~\ref{tab:overlap} indicate an improvement in performance when reducing the stride. From the original half-block overlap filter, a maximum improvement of \textbf{+\,0.42\,dB} is made at \sfrac{1}{7} of a block stride. We choose to keep a quarter-block overlap for our implementation as a finer overlap returns a performance gain too small for the gain in computational complexity.

\begin{table}[t]
\centering
\begin{tabular}{@{}lllllll@{}}
\toprule
Block Stride & 1/2   & 1/3   & \underline{1/4}   & 1/5   & 1/6   & 1/7   \\ \midrule
PSNR         & 35.06 & 35.33 & \underline{35.42} & 35.46 & 35.47 & 35.48 \\ \bottomrule
\end{tabular}
\caption{Block Overlap. Results are obtained using Hamming win-
dows for a 38\texttimes38 block with 1/x block overlaps. The 1/4 block overlap yields the best
compromise between gains (+\,0.36\,dB) and computation complexity.}
\label{tab:overlap}
\end{table}

\begin{figure*}[t]
\centering
\renewcommand{\arraystretch}{0.15}
\setlength{\tabcolsep}{0.15pt}
\begin{tabular}{cccccccccc}
    \includegraphics[width=0.095\textwidth]{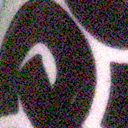} &    
    \includegraphics[width=0.095\textwidth]{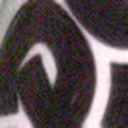} &
    \includegraphics[width=0.095\textwidth]{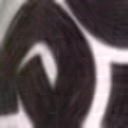}&
    \includegraphics[width=0.095\textwidth]{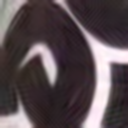}&
    \includegraphics[width=0.095\textwidth]{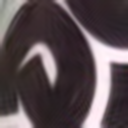}&
    \includegraphics[width=0.095\textwidth]{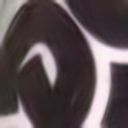}&
    \includegraphics[width=0.095\textwidth]{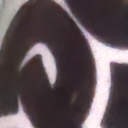}&
    \includegraphics[width=0.095\textwidth]{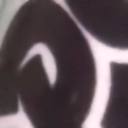}&
    \includegraphics[width=0.095\textwidth]{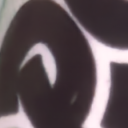}&
    \includegraphics[width=0.095\textwidth]{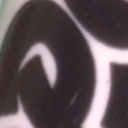}\\
    
    \includegraphics[width=0.095\textwidth]{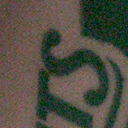} &
    \includegraphics[width=0.095\textwidth]{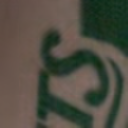}&
    \includegraphics[width=0.095\textwidth]{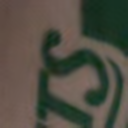}&
    \includegraphics[width=0.095\textwidth]{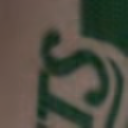} &
    \includegraphics[width=0.095\textwidth]{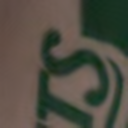}&
    \includegraphics[width=0.095\textwidth]{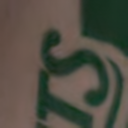}&
    \includegraphics[width=0.095\textwidth]{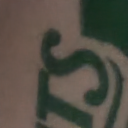}&
    \includegraphics[width=0.095\textwidth]{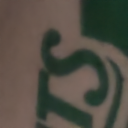}&
    \includegraphics[width=0.095\textwidth]{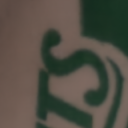}&
    \includegraphics[width=0.095\textwidth]{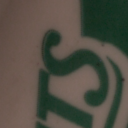}\\
    \includegraphics[width=0.095\textwidth]{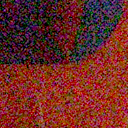} &
    \includegraphics[width=0.095\textwidth]{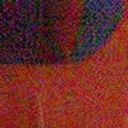} &
    \includegraphics[width=0.095\textwidth]{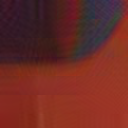}&
    \includegraphics[width=0.095\textwidth]{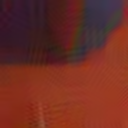}&
    \includegraphics[width=0.095\textwidth]{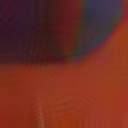}&
    \includegraphics[width=0.095\textwidth]{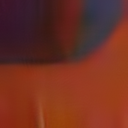}&
    \includegraphics[width=0.095\textwidth]{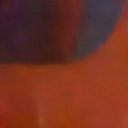}&
    \includegraphics[width=0.095\textwidth]{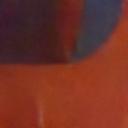}&
    \includegraphics[width=0.095\textwidth]{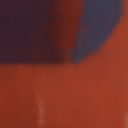}&
    \includegraphics[width=0.095\textwidth]{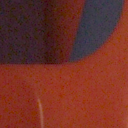}\\
    \includegraphics[width=0.095\textwidth]{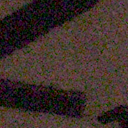}&
    \includegraphics[width=0.095\textwidth]{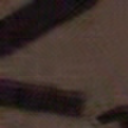}& 
    \includegraphics[width=0.095\textwidth]{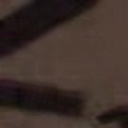}&
    \includegraphics[width=0.095\textwidth]{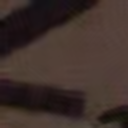}&
    \includegraphics[width=0.095\textwidth]{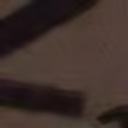}&
    \includegraphics[width=0.095\textwidth]{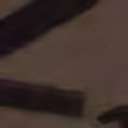}&
    \includegraphics[width=0.095\textwidth]{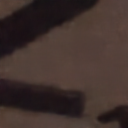}&
    \includegraphics[width=0.095\textwidth]{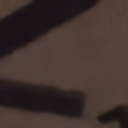}&
    \includegraphics[width=0.095\textwidth]{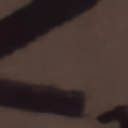}&
    \includegraphics[width=0.095\textwidth]{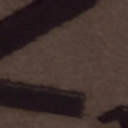}\\   

    \includegraphics[width=0.095\textwidth]{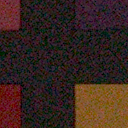}&
    \includegraphics[width=0.095\textwidth]{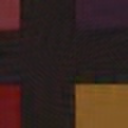}& 
    \includegraphics[width=0.095\textwidth]{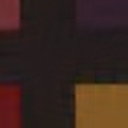}&
    \includegraphics[width=0.095\textwidth]{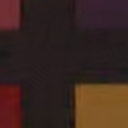}&
    \includegraphics[width=0.095\textwidth]{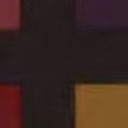}&
    \includegraphics[width=0.095\textwidth]{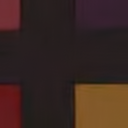}&
    \includegraphics[width=0.095\textwidth]{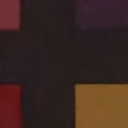}&
    \includegraphics[width=0.095\textwidth]{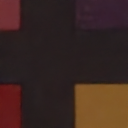}&
    \includegraphics[width=0.095\textwidth]{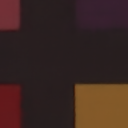}&
    \includegraphics[width=0.095\textwidth]{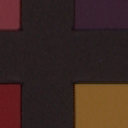}\\   
    (a) {noisy}\vphantom{\large{A}} & (b) W0 & (c) W1 & (d) W2 & (e) W3 & (f) W4 & (g) DnCNN & (g) CBDNet & (i) SUNet & (j) GT \\
    \end{tabular}
    \caption{Crop results for methods presented in Table~\ref{tab:summaryofresults}. W3 and W4 correspond to our optimised Wiener without and with NN coring.}
    \label{fig:res}
\end{figure*}

\vspace{0.25em}
\noindent\textbf{Analysis and Interpolation Windows.}  Setting the overlap to \sfrac{1}{4} of a window and the block size to 38\texttimes38, we found that replacing the half-cosine window with a Gaussian window with STD $\alpha=0.3$, improves our output PSNR to 35.52dB on our dataset, a \textbf{+\,0.1\,dB} improvement from the equivalent half-cosine window. 
Note that, as with most of these results, although improvements across the dataset are small, the improvement is still noticeable to the eye, with fewer blocky artefacts, as shown in figure~\ref{fig:res}.

\subsection{Per-Block Noise Estimation}\label{sec:stdnet}

\textbf{Training.} The ground-truth std maps are generated for 800 images of the DIV2k dataset~\cite{Agustsson_2017_CVPR_Workshops}, using the noise model of CBDNET~\cite{guo2019toward}. An additional 320 real-noise images from the SIDD training set~\cite{SIDD_2018_CVPR} are included for the fine-tuning stage.
Each of the 9 CNNs, described in section~\ref{sec:mean}, is trained for 4000 epochs with a cosine annealing learning rate which decays from \num{1e-3} to \num{1e-5} every 300 epochs. We use a mini-batch size of 24 and randomly select 128\texttimes128 image patches for every image.

\vspace{0.25em}
\noindent\textbf{Results.} In Table~\ref{tab:summary_std_table}, we first evaluate our 4 layer, 16 channel STD CNN against the Wiener filter with a baseline standard deviation estimation of $\sigma=10$. The performance steadily increases when progressing from per-image to per-channel and finally per-block estimation. We also include  results when using larger networks. We see a \textbf{+\,1.2\,dB} (36.72\,dB) improvement over our previous model with our smallest network (2 layer, 16 channel) and a maximum denoising performance of 37.20\,dB (\textbf{+\,1.68\,dB}) with our largest network. We choose to use our 4-layer networks for future experiments due to a more appealing performance to parameter count ratio. 

\begin{table}[t]
\centering
\begin{tabular}{@{}lllr@{}}
\toprule
Noise STD Prediction                           & Scope & PSNR(dB) & \#params\\ \midrule
Fixed to $\sigma$ = 10                         & Global    & 35.61   & -  \\
CNN, 4 layers, 16 channels & Per Image & 36.72    & 5.6k  \\
CNN, 4 layers, 16 channels & Per Channel  & 36.91    & 5.6k  \\

CNN, 2 layers, 16 channels & Per Block    & 36.72   & 0.9k   \\
CNN, 2 layers, 32 channels & Per Block    & 36.69   & 1.8k   \\
CNN, 2 layers, 64 channels & Per Block    & 36.70   & 3.6k   \\
CNN, 4 layers, 16 channels & Per Block    & 37.07   & 5.6k   \\
\underline{CNN, 4 layers, 32 channels} & \underline{Per Block}    & \underline{37.11}    & \underline{20.4k}  \\
CNN, 4 layers, 64 channels & Per Block    & 37.11   & 77.0k   \\
CNN, 6 layers, 16 channels & Per Block    & 37.12   & 10.0k   \\
CNN, 6 layers, 32 channels & Per Block    & 37.12   & 39.0k   \\
CNN, 6 layers, 64 channels & Per Block    & 37.20   & 152.0k   \\ \bottomrule
\end{tabular}
\caption{Effect of the Noise STD prediction on Wiener denoising results.  Quarter-overlap Gaussian windows on 38\texttimes38 blocks are used. The underlined method corresponds to our best compromise choice.}
\label{tab:summary_std_table}
\end{table}

\subsection{Block Mean Prediction}
In Table~\ref{tab:DC}, we report the results for the different block mean estimation methods proposed in section~\ref{sec:mean}. Results are obtained from our current best Wiener configuration (ie. Gaussian window, \sfrac{1}{4} block overlap, 38\texttimes38 and 4\texttimes16 standard deviation network). Using the median filter yields a PSNR of 37.55\,dB (\textbf{+\,0.48\,dB} over the mean). Use of other quantiles is discussed in the supplementary material.  

We also implement three custom U-Nets style networks to predict the mean values: UNet-L, UNet-M and UNet-S with 2M, 0.3M and 0.12M parameters respectively. Layer concatenations are replaced with summations to make relationships between skip connections easier to learn. The same training scheme is followed as the fine-tuning step section~\ref{sec:stdnet}. We obtain the results 37.24\,dB, 37.48\,dB and 37.41\,dB for the small medium and large networks respectively, and thus did not outperform the simple median filter.

Lastly, we fed the ground-truth image block means to the Wiener filter in an attempt to establish an upper bound of the potential gains that can be made by choosing better DC offsets. We record a benchmark result of 38.55\,dB, outlining the performance possible with further optimisation in that space.

\subsection{Multi-Scale Wiener Filtering}
Next, we evaluate the Wiener filters performance as an average of images filtered at different window sizes, specifically: 8\texttimes8, 16\texttimes16, 38\texttimes38, 64\texttimes64 and 96\texttimes96. We measured a \textbf{+\,0.32\,dB}  gain, with a practical best result of 37.87\,dB. Note that, with ground-truth values for the mean estimates, this performance could be raised to 38.82\,dB, showing that significant further gains should be attainable here. 


\begin{table}[t]
\centering
\setlength{\tabcolsep}{5pt}
\begin{tabular}{@{}lcccccc@{}}
\toprule
Method        & Mean      & Median & UNet-S & UNet-M & UNet-L & GT \\\midrule
PSNR & 37.07 & \underline{37.55} & 37.24 & 37.43 & 37.42 &  38.55\\
 \bottomrule
\end{tabular}
\caption{Comparison of Block Mean Estimation Methods. }
\label{tab:DC}
\end{table}

 \subsection{Coring Refinement Network}
Our final experiment explores the possibility of a frequency-CNN to fine-tune the derived Wiener coring function $H$ (see section~\ref{sec:coring_NN_method}). The network is trained using the same real-synthetic dataset as in our STD-CNN and using the same training parameters. For this experiment, we take the 4\texttimes32 STD-CNN predictor with frozen weights and the median DC-offset removal strategy. 
Our trained coring network yields an increased performance to 38.13\,dB, making a significant \textbf{+\,0.58\,dB} gain over the non-optimised coring result.

\subsection{Comparison with the State of the Art}

In Table~\ref{tab:summaryofresults} and Fig.~\ref{fig:res}, we compare our different optimisation levels of Wiener W0-4. W0 and W1 refer to the typical baseline Wiener scenarios, where $\sigma$ is either set arbitrarily or manually tuned on a per-image, per-channel basis. W2 refers to the improved windowing, with the use of median and per-block noise $\sigma$ estimation. W3 adds the quarter stride and multi-scale overlap, and W4 includes the coring network. These are compared to DnCNN~\cite{zhang2017beyond}, CBDNet~\cite{guo2019toward} and SUNet~\cite{liu2021swin}, a Swin-based transformer architecture.
We use the bias-free version of DnCNN~\cite{Mohan2020Robust}, and as it was originally trained for Gaussian noise, we re-trained it on our training set. 

Results in Table~\ref{tab:summaryofresults} suggest that W3 and W4 levels  compare favourably with the much larger DnCNN. Interestingly, we can see in Fig.~\ref{fig:res}-(c) that the coring network noticeably helps reduce the typical Wiener ringing artefacts.

\begin{table}[]\setlength{\tabcolsep}{5pt}
\begin{tabular}{@{}l@{}c@{}r@{}}
\toprule
Method                               & PSNR (dB) & ~~\#params  \\ \midrule
W0: Original Wiener & 35.06     & n/a            \\
\textit{W1: (W0) + per-channel $\sigma$} &  36.58   & n/a  \\
\textit{W2: (W0) + median + Gauss. + per-block $\sigma$} &  \textit{36.99}   & \textit{20k}  \\
\textit{W3: (W2) + \sfrac{1}{4} stride \& multi-scale overlaps}      & \textit{37.87}     & \textit{20k}       \\
DnCNN*                    & 37.94     & 640k           \\
\textit{W4: (W3) + Coring NN}        & \textit{38.17}     & \textit{43k}           \\
CBDNet                               & 39.32     & 4.36 M           \\
Swin/SUNet                           & 39.60     & 99.00 M            \\ \bottomrule
\end{tabular}
\caption{Comparison of some of our optimisation levels of Wiener on our real-noise benchmark. W3 and W4 optimisation levels are on par with a real-noise trained and bias-free DnCNN (DnCNN*). }
\label{tab:summaryofresults}
\end{table}
\section{Conclusions}\label{sec:conclusion}
We demonstrate that with careful consideration, the Wiener filter can perform on par, or even outperform popular CNNs such as DnCNN. We have put forward a novel method of automating STD selection using a small CNN, creating a blind Wiener denoiser. Artefacts such as ringing and blocking have been improved upon and the optimisation of window functions and finer overlapping allows for increased results. While we take advantage of neural networks to improve filtering performance, the final Wiener filter uses only 43k combined trainable parameters, while outperforming popular CNN-backbone denoisers which require millions of parameters. We note also that there is still much to gain from creating hybrid signal processing-CNN denoisers, adopting a best-of-both-worlds approach.

\bibliographystyle{IEEEbib}
\bibliography{strings,refs}

\end{document}